\documentclass[psamsfonts]{amsart}

\usepackage[foot]{amsaddr}
\usepackage{amssymb,amsfonts}
\usepackage[all,arc]{xy}
\usepackage{enumerate}
\usepackage{mathrsfs}
\usepackage{physics}
\usepackage{braket}
\usepackage{hyperref}
\usepackage[capitalize]{cleveref}
\usepackage{comment}

\newtheorem{thm}{Theorem}

\newtheorem{cor}[thm]{Corollary}

\newtheorem{lem}[thm]{Lemma}

\theoremstyle{definition}
\newtheorem{defn}{Definition}

\newtheorem{res}{Result}

\theoremstyle{remark}
\newtheorem{rem}[thm]{Remark}

\newtheorem*{Proof}{\bf Proof}

\title{Improved gap dependence in adiabatic state preparation by adaptive schedule}

\author[Xi Guo]{Xi Guo}
\address[Xi Guo]{Courant Institute of Mathematical Sciences, New York University, New York, 10012, USA; School of Mathematics Science, Peking University, Beijing, 100871, China}

\author[Dong An]{Dong An}
\address[Dong An]{Beijing International Center for Mathematical Research, Peking University, Beijing, 100871, China}
\email{dongan@pku.edu.cn}

\begin{document}

\begin{abstract}

Adiabatic quantum computing is a powerful framework for state preparation, while its evolution time often scales quadratically in the inverse Hamiltonian spectral gap, leading to sub-optimal computational complexity. 
In this work, we introduce a nonlinear adaptive strategy for finding the time scheduling function, and show that the gap dependence can be quadratically improved to be inverse linear for a wide range of systems under a mild gap measure condition. 
Through variational analysis, we further demonstrate the optimality of our schedule for systems with linear gap and the partial optimality for general systems, while we also rigorously show that the commonly used linear schedule is never optimal. 

\end{abstract}

\maketitle

\tableofcontents

\section{Introduction}

Preparing ground states of Hamiltonians appears ubiquitously across various areas of quantum physics, quantum chemistry, and quantum information science. 
Because standard algorithms executed on classical computers typically incur prohibitively expensive costs for large physical systems, researchers have been driven to investigate quantum approaches that might circumvent this bottleneck. 
Among many techniques, adiabatic quantum computing (AQC)~\cite{AlbashLidar2018} is a powerful and generic framework for quantum ground state preparation. 
In AQC, we consider a rescaled time-dependent Schr\"odinger equation 
\begin{equation}
    \frac{1}{T}i\frac{\partial}{\partial s}\ket{\psi_T(s)} = H(u(s))\ket{\psi_T(s)}. 
    \label{eq:schrodinger_intro}
\end{equation}
Here $s \in [0,1]$ is the rescaled time, and $T$ is the physical evolution time. 
The time-dependent Hamiltonian is defined as 
\begin{equation}
    H(u(s)) = (1-u(s))H_0 + u(s) H_1,
\end{equation}
which interpolates two time-independent Hamiltonians $H_0$ and $H_1$ through a smooth scheduling function $u(s)$ with $u(0) = 0$ and $u(1) = 1$. 
Then, under the gap condition of $H(u(s))$, if we initialize the dynamics to be an eigenstate of $H_0$, then the final solution $\ket{\psi_T(1)}$ for sufficiently large $T$ will be an approximation of the eigenstate of $H_1$. 
This yields an effective approach for ground state preparation of the Hamiltonian $H_1$, by first constructing a simple Hamiltonian $H_0$ with easily constructable eigenstate and simulating the dynamics with effective Hamiltonian simulation algorithms~\cite{HuyghebaertDeRaedt1990,WiebeBerryHoyerEtAl2010,PoulinQarrySommaEtAl2011,WeckerHastingsWiebeEtAl2015,LowWiebe2019,BerryChildsSuEtAl2020,AnFangLin2021,Yi2021,Kocia2022digital,AnCostaBerry2025,LuHuangAnEtAl2025}. 

A theoretical backbone of reliable adiabatic computation is the quantum adiabatic theorems (QATs), which quantify how well the actual quantum evolution of~\cref{eq:schrodinger_intro} can approximate the ideal ground state. 
Since the pioneering works of Born and Fock \cite{BornFock1928} and Kato \cite{Kato1950}, there have been considerable subsequent works proposing improved versions of QATs~\cite{Nenciu1993,AvronElgart1999,HagedornJoye2002,AmbainisRegev2006,JansenRuskaiSeiler2007,LidarRezakhaniHamma2009,CheungHoyerWiebe2011,ElgartHagedorn2012,GeMolnarCirac2016,MozgunovLidar2022}. 
Among them, Jansen et al.\cite{JansenRuskaiSeiler2007} proposed an explicit error formula under weak smoothness assumptions of Hamiltonians, explicitly quantifying the relationship between the Hamiltonian's spectral gap $\Delta(s)$ and evolution time $T$ and significantly expanding the theory's applicability.

Specifically, the QATs reveal that in generic adiabatic dynamics, the evolution time $T$ should be chosen as $\mathcal{O}(1/\Delta_*^2)$ to ensure a sufficiently accurate adiabatic approximation, where $\Delta_*$ denotes the minimal spectral gap of $H(u(s))$ over the entire time interval. 
Though explicit and generic, such a quadratic dependence on the inverse gap can often lead to sub-optimal scaling of quantum algorithms based on AQC. 
For example, in the adiabatic Grover search case~\cite{RolandCerf2002}, the minimal gap $\Delta_* \sim 1/\sqrt{N}$ where $N$ is the dimension of the Hilbert space, so a naive implementation of AQC needs to evolve the quantum dynamics up to $T \sim N$, ruining the quadratic quantum speedup. 
Another example is the adiabatic-based quantum linear system algorithms~\cite{SubasiSommaOrsucci2019,AnLin2022}, where the minimal gap scales linearly in the condition number $\kappa$ of the matrix. 
As a result, the computational cost of a vanilla AQC algorithm scales quadratically in terms of the condition number, which is again a sub-optimal scaling. 
Therefore, it is desirable and practically relevant to explore whether or not the gap dependence of the AQC simulation time could be further improved. 

In this work, we introduce a nonlinear scheduling strategy based on gap-dependent parametrization. 
Through power-law scheduling $u'(s)\propto\Delta^p(u(s))$ motivated by~\cite{RolandCerf2002,SubasiSommaOrsucci2019,AnLin2022}, where $\Delta(u(s))$ is the spectral gap of $H(u(s))$, the system adaptively modulates its evolution speed, decelerating in small-gap regions while accelerating through large-gap phases. 
This ``slow-in-critical-zones'' strategy achieves dual synergistic effects: 
suppressing error accumulation in bottleneck regions while maintaining computational efficiency elsewhere.
We prove that this scheduling reduces the adiabatic error scaling to $\mathcal{O}(\Delta_*^{-1})$ for a wide range of systems under a mild gap measure condition, achieving quadratically improved gap dependence. 
Furthermore, through variational analysis, we demonstrate the optimality of the power-law scheduling with $p=3/2$ for linear gap profiles as well as its partial optimality for general systems, establishing theoretical foundations for practical implementations.

\subsection{Main results}

The main technique to improve the gap dependence is an adaptive power-law scheduling. 
Specifically, we choose the scheduling function to be the solution of the ordinary differential equation (ODE)
\begin{equation}
    u'(s) \propto \Delta^p(u(s)), \quad s \in [0,1], \quad u(0) = 0, u(1) = 1. 
    \label{eq:power_law_intro}
\end{equation}
Here $\Delta(u(s))$ denotes the spectral gap of the time-dependent Hamiltonian $H(u(s))$, and $p \in (1,2)$ is a fixed parameter. 
Sometimes we also use the notation $\Delta(u)$ to emphasize its intrinsic dependence on the scheduling rather than the time $s$. 
Without loss of generality, we assume that the norms of the Hamiltonians $H_0$ and $H_1$ are bounded by $1$, otherwise we may rescale the Hamiltonian and absorb the norm in the spectral gap. 

\subsubsection{Quadratically improved error bound}

Our first main result is an improved adiabatic error bound when using this power-law scheduling. 

\begin{res}[Informal version of \Cref{thm:error_estimation}]\label{res:error_bound}
        When the gap $\Delta(u)$ satisfies the measure condition that $ \mu\Big(\big\{ u \in [0,1] \,:\, \Delta(u) \leq x \big\}\Big) = \mathcal{O}(x)$ with Lebesgue measure $\mu$, the adiabatic error with the power-law scheduling in \cref{eq:power_law_intro} becomes $\mathcal{O}(T^{-1}\Delta_*^{-1})$. 
\end{res}

\cref{res:error_bound} implies that it suffices to choose $T \sim 1/\Delta_*$ for constant accuracy, which is a quadratic improvement compared to using a naive linear schedule. 
The proof of \cref{res:error_bound} leverages the general QAT in~\cite{JansenRuskaiSeiler2007} with a tailored analysis for the power-law schedule. 
A key condition in the proof as well as for the quadratic speedup is the measure condition, which has been imposed in an existing work~\cite{JarretLackeyLiuWan2019}. 
The intuition behind the measure condition is that the spectral gap does not remain small for too long, preventing dense accumulations of small-gap regime which might significantly amplify adiabatic errors. 
Notice that this measure condition holds for important applications such as the adiabatic Grover search and the adiabatic quantum linear system algorithms, as well as a wide range of Hamiltonians where there are finitely many locally minimal spectral gaps along the trajectory. 

As long as the measure condition holds, the quadratic speedup claimed in~\cref{res:error_bound} can be achieved without any further requirements on the Hamiltonians. 
We would like to remark that implementing the power-law scheduling requires a priori knowledge on the spectral gap, which is a big assumption involving another QMA-hard computational task. 
However, fortunately, there exist useful cases, such as adiabatic Grover search, 2-SAT on a ring problem and adiabatic quantum linear system algorithms, where at least a constantly accurate lower bound of the spectral gap is known. 
In this scenario, the power-law scheduling can be efficiently constructed either by analytically solving the ODE in~\cref{eq:power_law_intro}, yielding a closed-form representation, or by high-precision numerical ODE solvers, which only require the gap information over logarithmically many discrete time points and do not explicitly depend on the dimension of the Hamiltonians.

\subsubsection{Variational optimality analysis}

Our second main result is an optimality analysis of the power-law schedule. 
Our starting point is the adiabatic error bound established in~\cite{JansenRuskaiSeiler2007}, which contains two components: $\int_0^1 ( \|H^{(1)}(s)\|^2 / \Delta^3(u(s)) ) ds$, the term with the first-order time derivative of the Hamiltonian $H(u(s))$; and $\int_0^1 ( \|H^{(2)}(s)\| / \Delta^2(u(s)) ) ds$, the term with the second-order derivative. 

\begin{res}[Informal version of \Cref{thm:L2_optimal} and \cref{thm:opt_lgap}]\label{res:optimality}
    The power-law schedule defined in~\cref{eq:power_law_intro} satisfies: 
    \begin{enumerate}
        \item When the gap $\Delta(u)$ is linear in $u$, it minimizes the adiabatic error bound of~\cite{JansenRuskaiSeiler2007}. 
        \item For the general-form gap, it minimizes the term with the first-order time derivative in the adiabatic error bound of~\cite{JansenRuskaiSeiler2007}. 
    \end{enumerate}
\end{res}

\cref{res:optimality} implies that the power-law schedule with $p=3/2$ is optimal when the gap is linear. 
Here the ``optimality'' is in the sense that the schedule minimizes the adiabatic error upper bound in~\cite{JansenRuskaiSeiler2007}. 
We note that this does not mean that the actual adiabatic error is minimized, because the upper bound in~\cite{JansenRuskaiSeiler2007} could possibly yield an imprecise description of the dynamics by overestimating the actual adiabatic error. 
Nevertheless, our focus is on conducting an optimality analysis with rigorous performance guarantee, so we choose to investigate the upper bound in QATs. 

When the gap is no longer linear, \cref{res:optimality} shows that the power-law schedule with $p = 3/2$ is partially optimal, minimizing the first-order time derivative term in the adiabatic error bound. 
Notice that the first-order time derivative term involves a cubic inverse gap dependence, while the other second-order time derivative term only exhibits a quadratic inverse dependence. 
So we expect that in practice, the first-order time derivative term is likely to be dominant, and the power-law schedule with $p = 3/2$ is still practically preferable by optimizing the majority of the adiabatic error bound. 

Our analysis is based on the variational method for the error functional defined through the adiabatic error bound. 
For this functional, we compute its Euler-Lagrange equation, which is the first-order optimality condition, and use this to study the optimality of the power-law schedule. 
As a side product, we also rigorously show a common intuition that the linear schedule (i.e., $u(s) = s$) does not satisfy the Euler-Lagrange equation unless the spectral gap is trivially constant. 
This necessitates the usage of nonlinear schedule functions. 

\begin{res}[Informal version of \Cref{thm:nonconstant_gap}]\label{res:constant}
    Once the spectral gap $\Delta(u(s))$ is not a constant function, the linear schedule $u(s) = s$ is not optimal, i.e., there exists a nonlinear schedule with smaller adiabatic error bound. 
\end{res}

\subsection{Related works}

As the gap dependence is crucial to the efficiency of AQC, numerous existing works have been done to improve the gap dependence in AQC by optimizing the schedule~\cite{RezakhaniKuoHammaEtAl2009,RezakhaniAbastoLidarZanardi2010,JarretLackeyLiuWan2019,Isermann2021,MatsuuraBuckSenicourtZaribafiyan2021,WanKim2022,ShinguHatomura2025,BraidaChakrabortyChaudhuriEtAl2025}. 
For example, the quantum adiabatic brachistochrone framework~\cite{RezakhaniKuoHammaEtAl2009} finds the schedule that minimizes a geometric functional related to the approximate version of the adiabatic error, and similar ideas have been further explored~\cite{RezakhaniAbastoLidarZanardi2010,Isermann2021}. 
These works require knowledge of the spectral gap to construct the schedule. 
Recently, the so-called gap-uninformed strategies have also been proposed based on several different techniques such as the variational framework~\cite{MatsuuraBuckSenicourtZaribafiyan2021}, Krylov subspace approximation to the adiabatic gauge potential~\cite{ShinguHatomura2025}, and time evolution with constant speed schedule~\cite{HanParkChoi2025}.  
Our work contributes to this topic by rigorously establishing the quadratic gap improvement for a wide range of Hamiltonians, giving an alternative yet simple strategy for constructing the corresponding schedule when the gap is known, and demonstrating the optimality of this schedule rigorously as well.

The key component of our work is the power-law schedule defined through an ODE in~\cref{eq:power_law_intro}. 
The power-law schedule has already been proposed and explored previously for several specific applications. 
For example, the case with $p = 2$ was proposed for the adiabatic Grover search algorithm in~\cite{RolandCerf2002} to achieve quadratic speedup, and recently the case $1 \leq p \leq 2$ has been further explored for more general unstructured adiabatic quantum optimization problem~\cite{BraidaChakrabortyChaudhuriEtAl2025}. 
In the adiabatic-based quantum linear system algorithms~\cite{SubasiSommaOrsucci2019,AnLin2022} with near-optimal query complexity, the schedule is also the power-law schedule with $1 < p < 2$. 
Our work extends the scope by showing the improved gap dependence brought by the power-law schedule for general systems with measure condition. 
We remark that the bashful adiabatic algorithm proposed in~\cite{JarretLackeyLiuWan2019} also naturally produces an adiabatic path with approximately the power-law schedule of $p = 1$, while our optimality analysis yields the significance of choosing $p$ to be larger than $1$.

\subsection{Organization}

The rest of paper is organized as follows. 
Section 2 introduces the quantum adiabatic theorem and error bounds under general scheduling. 
Section 3 enhances error bounds through spectral gap measure conditions and power-law scheduling. 
Section 4 analyzes scheduling optimality via variational methods. 
Section 5 presents concluding discussions.

\section{Preliminary}

\subsection{Set up}

Assume the Hamiltonian evolves as
\begin{equation}
    H(u(s)) = \left(1 - u(s)\right)H_0 + u(s)H_1, \quad 0 \leq s \leq 1,
\end{equation}
where $H_0$ and $H_1$ are time-independent Hamiltonians and the function $u$: $[0,1]\to[0,1]$ is called scheduling function, which is a strictly increasing mapping with $u(0)=0$ and $u(1)=1$.
Consider the adiabatic evolution
\begin{equation}
    \frac{1}{T}i\frac{\partial}{\partial s}\ket{\psi_T(s)} = H(u(s))\ket{\psi_T(s)},
    \label{eq:schrodinger1}
\end{equation}
where $0\leq s\leq 1$, and the parameter $T$ is called runtime of AQC. Here $\ket{\psi_T(s)}$ describes the adiabatic evolution with initial condition $\ket{\psi_T(0)}$ in the ground state of $H(0)=H_0$.
Let \( P_0(s) \) denote the projection operator onto the instantaneous ground state space of \( H(u(s)) \), and \( \Delta(u(s)) \) represent the energy difference between the ground state and the first excited state of $H(u(s))$.

With the projection operator $P_0$ and the gap function, we will provide a more precise characterization of the error in adiabatic evolution in the following QATs.

\subsection{Quantum Adiabatic Theorems (QATs)}

For static Hamiltonians (\( H(u(s)) \equiv H \)), the solution simplifies to \( \ket{\psi_T(s)} = \exp(-iHTs)\ket{\psi_0(0)} \), preserving the ground state during evolution.
However, if the Hamiltonian varies with time, we can no longer directly assert that the evolution remains within the instantaneous eigenspace. 
In this case, the evolution exhibits complex dynamical behavior, necessitating further detailed analysis, and the quantum adiabatic theorems give a formal description of this phenomenon.

\begin{lem}[QAT~\cite{JansenRuskaiSeiler2007}]
\label{thm:general_schedule}
The adiabatic approximation error satisfies
\begin{equation}
    \left\lvert 1 - \Braket{{\psi}_T(s)| P_0(s) | {\psi}_T(s)} \right\rvert \leq \eta^2(s),
\end{equation}
where
\begin{equation}
\label{eq:general_schedule_error}
    \eta(s) 
    = 
    \frac{C}{T}
    \left\{
    \frac{\|H^{(1)}(0)\|_2}{\Delta^2(0)} 
    + 
    \frac{\|H^{(1)}(s)\|_2}{\Delta^2(u(s))} 
    + 
    \int_{0}^{s} 
    \left(
    \frac{\|H^{(2)}(s')\|_2}{\Delta^2(u(s'))} 
    + 
    \frac{\|H^{(1)}(s')\|_2^2}{\Delta^3(u(s'))}
    \right) \mathrm{d}s'
    \right\}.
\end{equation}
Here $H^{(k)}(s) = \frac{\mathrm{d}^k}{\mathrm{d}s^k}H(u(s))=\left(H_1-H_0\right)\frac{\mathrm{d}^k}{\mathrm{d}s^k}u(s)$ denotes the $k$-th derivative with respect to $s$ for $k = 1, 2$, and $C$ is a constant independent of $s$, $\Delta$, and $T$.
\end{lem}

Its fundamental premise states that if a quantum system starts in the ground state of an initial Hamiltonian \( H_0 \), and the Hamiltonian evolves sufficiently slowly to a final Hamiltonian \( H_1 \), the system remains near the instantaneous ground state throughout the process.

The introduction of the time evolution interpolation function in \Cref{thm:general_schedule} explicitly reveals how the system’s evolution rate, captured by the terms $du/ds$ and $d^2u/ds^2$, affects the upper bound of the error, since the approximation accuracy depends critically on two competing factors: the spectral gap $\Delta(u(s))$ and the Hamiltonian derivatives $\|H^{(1)}(s)\|_2$ and $\|H^{(2)}(s)\|_2$.
This observation suggests that by appropriately controlling the rate of evolution, we may reduce the error more effectively. 
We will discuss this in detail in later section. Before that, let us first consider what the error would look like if no scheduling function were introduced. In other words, we may consider the simplest choice of scheduling function $u(s)=s$. As a consequence of \Cref{thm:general_schedule}, the following corollary provides an upper bound on the error associated with linear interpolation.

\begin{cor}[QATs for linear scheduling function]\,\\
\label{cor:2}
\noindent With the linear schedule $u(s)=s$, the adiabatic approximation error satisfies:
\begin{equation}
    \left\lvert 1 - \Braket{\psi_T(s)| P_0(s) | \psi_T(s)} \right\rvert \leq \eta^2(s),
\end{equation}
where
\begin{equation}
\label{eq:linear_error_co2}
    \eta(s) 
    = 
    \frac{C}{T}
    \left\{
    \frac{\|H'(0)\|_2}{\Delta^2(0)}
    + 
    \frac{\|H'(s)\|_2}{\Delta^2(s)}
    + 
    \int_{0}^{s} 
    \frac{\|H'(s')\|_2^2}{\Delta^3(s')}
    \mathrm{d}s'
    \right\}.
\end{equation}
Here $H'(s)=\frac{d}{ds}H(s)$.
$C$ is a constant independent of $s$, $\Delta$, and $T$.
\end{cor}

\section{The improvement of error bound}
In the previous section, \Cref{thm:general_schedule} suggested that it is possible to design an appropriate evolution strategy for the system by accelerating or decelerating the evolution in a controlled manner to reduce the error. 
In this section, we formalize this intuition. 

We begin by imposing certain constraints, referred to as measure conditions, on the spectral gap function. 
For spectral gap functions that satisfy these conditions, we will then introduce a power-law condition on the time scheduling function. 
And finally, we will prove that scheduling functions satisfying this power-law condition can further reduce the error to $\mathcal{O}\left(\Delta^{-1}_*\right)$.

\subsection{Measure condition for spectral gap}
The adiabatic evolution's fidelity crucially depends on the spectral gap's behavior throughout the parameterization. 
To quantify pathological scenarios where small gaps might accumulate, we introduce a measure-criterion constraining the prevalence of near-degeneracies.

\begin{defn}[Spectral gap measure condition]
\label{def:measure_condition}\,

\noindent The spectral gap function $\Delta: [0,1] \to \mathbb{R}^+$ of Hamiltonian\footnote{Throughout this work, the general time-dependent Hamiltonian we consider is denoted as $H(u(s))$ to highlight the dependence on the schedule function $u(s)$. We use $H(s)$ to denote the linearly interpolated Hamiltonian $H(s) = (1-s)H_0 + sH_1$. These notations are consistent because we can choose $u(s) = s$ for linear interpolation. Similar notations are used for the gap $\Delta(u(s))$. } $H(s) = (1 - s)H_0 + sH_1$ satisfies the \textbf{measure condition} if there exists a constant $C > 0$, such that for all $x > 0$, 
\begin{equation}
    \mu\Big(\big\{ s \in [0,1] \,:\, \Delta(s) \leq x \big\}\Big) \leq Cx,
    \label{eq:measure_cond}
\end{equation}
where $\mu$ denotes the Lebesgue measure on the interval $[0,1]$.
\end{defn}

This condition enforces sparse distribution of small spectral gaps.
The measure of parameters $s$ where $\Delta(s)$ drops below $x$ decays at least linearly with $x$. 
In other words, the total duration during which the spectral gap is smaller than $x$ scales at most linearly with $x$; the system cannot remain in a regime of small spectral gap for too long and must obey a linear upper bound.
Such control prevents dense accumulations of near-degenerate configurations that could catastrophically amplify adiabatic errors.

With the assumption of $\Delta(s)\geq\Delta_*>0$, we know that no level crossings occur during the evolution,  and Kato’s spectral perturbation theory implies that the gap function is smooth. 
The analyticity of $\Delta(s)$ admits a piecewise linear lower bound $\Delta_l(s)$: we can first partitions $[0,1]$ into sufficiently small subintervals such that the convexity or concavity of $\Delta(s)$ does not change within each subinterval. 
On intervals where $\Delta(s)$ is convex, a lower bound is given by the tangent line, while on intervals where it is concave, a lower bound is given by the secant line. 
In this way, one constructs a piecewise linear function $\Delta_l(s)$ such that, in the small-gap regime, its sublevel sets consist of finitely many V-shaped valleys, with slopes' absolute value bounded below by $m_* > 0$. 
Under these conditions, the measure condition holds with a constant $C$ depending on the number of valleys N and the minimal slope $m_*$, e.g., $C\approx 2N/m_*$.

Notably, many Hamiltonian operators satisfy this condition in quantum systems, including:

\begin{itemize}
    \item \textbf{Adiabatic quantum linear system solver~\cite{AnLin2022}}: 
    For $H(s)= (1 - s)H_0 + sH_1$, where $H_0=\sigma_x\otimes Q_b$ and $H_1=\sigma_+\otimes (AQ_b)+\sigma_-\otimes (Q_bA)$, we have $\Delta(s)\geq 1-s+s/\kappa$, showing that the gap satisfies the measure condition with $C=\frac{\kappa}{\kappa-1}$, where $\kappa=\|A\|\|A^{-1}\|$ is the condition number of $A$.
    
    \item \textbf{Adiabatic quantum Grover algorithm ~\cite{RolandCerf2002}}: 
    The gap function 
    $$
    \Delta(s)=\sqrt{(1-2s)^2+\frac{4}{N}s(1-s)},
    $$
    satisfies the measure condition with $C=\sqrt{\frac{N}{N-1}}$.
\end{itemize}
These systems collectively demonstrate the proposed measure condition's validity and broad applicability.

\subsection{Power-law scheduling function}
\label{subsec:power_scheduling}

The scheduling function's parametrization directly determines the rate of system evolution and thus crucially impacts adiabatic error scaling. 
In the following, we introduce a power-law parametrization strategy that adapts the schedule to spectral gap variations by adaptively adjusting the interpolation function in response to variations in the spectral gap.

\begin{defn}[Power-law scheduling condition]
\label{def:power_condition}\,

\noindent A scheduling function $u: [0,1] \to [0,1]$ with boundary conditions $u(0)=0$, $u(1)=1$ is said satisfying power condition with exponent $p \in (1,2)$ if its derivative satisfies:
\begin{equation}
    u'(s) = c_p \Delta^p(u(s)), \quad s \in [0,1],
    \label{eq:power_law}
\end{equation}
where $\Delta(u(s))$ is the spectral gap at parameter $u(s)$, and $c_p=\int_0^1 \Delta^{-p}(u) du $ is the normalization constant determined through integration of the gap structure and ensuring proper boundary conditions.
\end{defn}

From the power-law condition \cref{eq:power_law}), we observe that when $\Delta(u)$ is small, the instantaneous evolution rate $u'$ is correspondingly reduced; whereas when $\Delta(u)$ is large $u'$ increases. 
This implies that the time interpolation function satisfying $\frac{du}{ds} \propto \Delta^{p}(u)$ serves as an adaptive mechanism for regulating the evolution rate — slowing down the process in regions with small spectral gaps and accelerating it where the gap is large.

Moreover, to satisfy the boundary condition $u(1) = 1$ under the initial condition $u(0) = 0$, a normalization constant $c_p$ is required, which is given by the following expression:
\begin{equation}
    c_p
    =
    \int_{u^{-1}(0)}^{u^{-1}(1)}c_p\;\mathrm{d}s
    =
    \int_{u^{-1}(0)}^{u^{-1}(1)} \Delta^{-p}(u(s))\cdot c_p\Delta^{p}(u(s))\;\mathrm{d}s
    =
    \int_0^1 \Delta^{-p}(u)\;\mathrm{d}u.
    \label{eq:c_p}
\end{equation}
Intuitively, when an interpolation function satisfying the power-law condition is employed, the numerator of the integrands in the error bound \cref{eq:general_schedule_error} from  \Cref{thm:general_schedule} will involve powers of $\Delta(u)$. 
As a result, the effective order of the denominators $\Delta^2(u)$ and $\Delta^3(u)$ is reduced, thereby weakening the singular behavior associated with small spectral gaps.

In the next subsection, we will see that this interpolation strategy allows the system to slow down its evolution in regions where the spectral gap approaches zero, thus enabling improvements in error precision and potentially reducing the overall error scaling.

\subsection{Error bound enhancement via schedule optimization}
In this section, we analyze the dependence of the adiabatic error on the minimum gap $\Delta_*$. Throughout the remaining part of this paper, we assume that $\Delta(s)$ satisfies the measure condition. 
We will first give the error scaling in the absence of a scheduling function, as stated in the theorem below.
\begin{thm}
\label{thm:error_estimation_wo_sf}
Let the spectral function $\Delta(s)$ of $H(s)=(1-s)H_0+sH_1$ satisfies measure condition with constant $C$, then the total adiabatic error in \Cref{eq:linear_error_co2} at $s=1$ satisfies:
\begin{equation}
    \eta(1)\leq \mathcal{O}\left(\Delta^{-2}_*\right).
    \label{eq:enhanced_bound}
\end{equation}
where the minimal gap is defined as:
\begin{equation}
    \Delta_* = \inf_{s \in [0,1]} \Delta(u(s)) > 0.
    \label{eq:min_gap}
\end{equation}
\end{thm}
To establish the proof of this theorem, we require the following estimate controlling the gap integral.

\begin{lem}[Spectral gap integral bound]
    \label{lem:gap_integral}\,\\
    \noindent Under the measure condition \cref{eq:measure_cond}, the gap-weighted integral satisfies
    \begin{equation}
    \int_0^1 \Delta^{-\alpha}(u) du \leq C_\alpha \Delta_*^{-(\alpha - 1)}
    \label{eq:contral_lemma}
    \end{equation}
    for any exponent \(\alpha > 1\). Here $C_\alpha$ is a positive constant which only depends on $C$ and $\alpha$.
\end{lem}

\noindent \textbf{Proof of \Cref{lem:gap_integral}.} The analysis begins with transforming this gap-weighted integral into its equivalent form of measure:
\[
\int_0^1 \Delta^{-\alpha}(u) du = \int_0^\infty \mu\left(\{u : \Delta^{-\alpha}(u) > t\}\right) dt = \int_0^\infty \mu\left(\{u : \Delta(u) < t^{-1/\alpha}\}\right) dt,
\]
where $\mu$ denotes Lebesgue measure and the equality follows from the monotonic decay of $\Delta^{-\alpha}(u)$ with respect to $\Delta(u)$. 
Introducing $t_0 = \Delta_*^{-\alpha}$, we observe that $\Delta(u) \geq \Delta_* > t^{-1/\alpha}$ holds for all $t \leq t_0$, making the set $\{u : \Delta(u) < t^{-1/\alpha}\}$ a null set. 
This allows restriction of the integration domain to $t > t_0$:
\[
\int_{t_0}^\infty \mu\left(\{u : \Delta(u) < t^{-1/\alpha}\}\right) dt.
\]
Utilizing the measure condition $\mu(\{u : \Delta(u) < x\}) \leq Cx$ from \cref{eq:measure_cond} with $x = t^{-1/\alpha}$, we bound the integral by
\[
\int_{t_1}^\infty C t^{-1/\alpha} dt = C_\alpha\Delta_*^{-(\alpha-1)},
\]
where $C_\alpha=\frac{C\cdot\alpha}{\alpha-1}$, demonstrating how the minimal gap $\Delta_*$ governs the integral.\qed

\vspace{0.8em}

Having established the foundational estimate in \Cref{lem:gap_integral}, we now apply this result to error bound. 

\vspace{0.8em}

\noindent \textbf{Proof of \Cref{thm:error_estimation_wo_sf}.}
We know $H(s)=(1-s)H_0+sH_1$ and hence $H'(s)=H_1-H_0$. Now we will proceed to examine each term in \Cref{eq:linear_error_co2} separately.

For the first two terms, we can readily obtain the following result.
$$
\frac{\|H'(s)\|_2}{\Delta^2(s)}
=\frac{\|H_1-H_0\|_2}{\Delta^2(s)}
\leq \|H_1-H_0\|_2\cdot \Delta_*^{-2}
=\mathcal{O}(\Delta_*^{-2}).
$$
The above expression holds for all $s \in [0,1]$. By substituting $s = 0$ and $s = 1$, we immediately obtain upper bounds for the first two terms.

Next, for the third term, we first apply \Cref{lem:gap_integral} with $\alpha = 3$ to obtain the following result,
$$
\int_0^1 \Delta^{-3}(u) du \leq \mathcal{O} \left(\Delta_*^{-2}\right)
$$
from which an upper bound for the third term follows directly as follow:
$$
\int_{0}^{1} 
\frac{\|H'(s)\|_2^2}{\Delta^3(s)}
\mathrm{d}s
=
\|H_1-H_0\|_2\cdot
\int_0^1 \Delta^{-3}(s) ds
\leq \mathcal{O} \left(\Delta_*^{-2}\right).
$$
This completes the proof of \Cref{thm:error_estimation_wo_sf}.\qed

\vspace{0.8em}

This theorem shows that, without the use of a scheduling function, the adiabatic error scales quadratically with respect to $\Delta_*$.
The following theorem provides an analysis of the adiabatic error under a power-law scheduling strategy, since the interplay between spectral measure conditions and power-law scheduling yields significant error suppression.
\begin{thm}
\label{thm:error_estimation}
Let the spectral function $\Delta(s)$ of $H(s)$ satisfies measure condition with constant $C$ and $u(s)$ satisfies power condition with exponent $p \in (1,2)$, then the total adiabatic error at $s=1$ with scheduling satisfies:
\begin{equation}
    \eta(1)\leq \mathcal{O}\left(\Delta^{-1}_*\right).
\end{equation}
where the minimal gap is defined as:
\begin{equation}
    \Delta_* = \inf_{s \in [0,1]} \Delta(u(s)) > 0.
\end{equation}
\end{thm}

\noindent \textbf{Proof of \Cref{thm:error_estimation}.}
Recall our notations:
\begin{align*}
    H^{(k)}(s) &= \frac{\mathrm{d^k}}{\mathrm{d}s^k}H(u(s))=\left(H_1-H_0\right)u^{(k)}(s),
\end{align*}
since the scheduling function $u(s)$ satisfies the power condition, we have
\begin{align*}
    u''(s)&=c_p\cdot p\Delta^{p-1}(u(s))\cdot\Delta'(u(s))\cdot u'(s)\\
    &=p\cdot c_p^2\cdot\Delta^{2p-1}(u(s))\cdot\Delta'(u(s)),
\end{align*}
then the total adiabatic error at $s = 1$ is
\begin{equation}
    \eta(1) 
    = 
    \frac{A\cdot C}{T}
    \left\{
    \frac{|u'(0)|}{\Delta^2(0)} 
    + 
    \frac{|u'(1)|}{\Delta^2(1)} 
    + 
    \int_{0}^{1} 
    \left(
    \frac{|u''(s')|}{\Delta^2(u(s'))} 
    + 
    \frac{A(u'(s))^2}{\Delta^3(u(s'))}
    \right) 
    \mathrm{d}s'
    \right\}.
    \label{eq:main_bound}
\end{equation}
where $A=\|H_1-H_0\|_2$ is the norm of $H_1-H_0$.

Prior to proving each term in \cref{eq:main_bound}, we first establish an upper bound for the normalization constant $c_p$. 
By using \Cref{lem:gap_integral} with $\alpha=p\in(1,2)$, we yield the upper bound of $c_p$:
\begin{equation}
    c_p \leq C_p\Delta_*^{-(p-1)},
    \label{eq:bound_cp}
\end{equation}
where $C_p=\frac{C\cdot p}{p-1}$, demonstrating how the minimal gap $\Delta_*$ governs the normalization constant.

Now, let us prove the four terms in the above equation separately.

For the first and second terms, also the boundary terms in the error estimation, let us begin with the power condition established in \cref{eq:power_law}, and then analyze these two terms:
\[
\frac{|u'(s)|}{\Delta^2(u(s))} = c_p \Delta^{p-2}(u(s)),
\]
where the right-hand side emerges from direct substitution of the power condition. 
Substituting the previously derived upper bound of $c_p$ and recognizing $\Delta(u(s)) \geq \Delta_*$ by definition of the minimal gap, we bound the ratio at the endpoints:
$$
\frac{|u'(0)|}{\Delta^2(0)}\leq C_p\Delta_*^{-1}
\quad
\text{and}
\quad
\frac{|u'(1)|}{\Delta^2(1)}\leq C_p\Delta_*^{-1}.
$$

The third error component involves the integral of the second derivative of scheduling function $u(s)$.
Beginning with the expression  
$$  
\frac{|u''(s')|}{\Delta^2(u(s'))} = p\cdot c_p^2 \cdot \Delta^{p-3}(u(s')) \cdot |\Delta'(u(s'))| \cdot u'(s'),  
$$  
we perform integration variable substitution $s' \mapsto u = u(s')$. 
This transforms the integral into
$$  
\int_0^1 \frac{|u''(s')|}{\Delta^2(u(s'))} ds' = p c_p \int_0^1 \Delta^{p-3}(u) |\Delta'(u)| du.  
$$

Before analyzing the bound for this component, we demonstrate that $\Delta'(u(s))$ is bounded by spectral perturbation theory.
Consider the parameterized Hamiltonian $H(u) = (1-u)H_0 + uH_1$ with Hermitian structure, whose eigenvalues $\lambda_i(u)$ satisfy non-decreasing ordering $\lambda_1(u) \leq \lambda_2(u) \leq \cdots \leq \lambda_n(u)$, then for an infinitesimal parameter perturbation $\delta u$, Weyl's inequality states that:
$$
|\lambda_i(u+\delta u) - \lambda_i(u)| \leq \|H(u+\delta u) - H(u)\|_2.
$$
It is trivial to see that $\|H(u+\delta u)-H(u)\|_2=\delta u\cdot A$.
Dividing both sides by $\delta u$ and taking the limit as $\delta u \to 0$ yields the bound for the derivative of eigenvalue:
$$
\left|\frac{d\lambda_i}{du}\right| \leq A, \quad \forall i=1,\,2,\,\cdots,\,n.
$$
Applying triangle inequality to the spectral gap $\Delta(u) = \lambda_2(u) - \lambda_1(u)$ yields:
$$
|\Delta'(u)| \leq \left|\frac{d\lambda_2}{du}\right| + \left|\frac{d\lambda_1}{du}\right| \leq 2A.
$$

Then, applying \Cref{lem:gap_integral} with exponent $\alpha = 3 - p \in (1,2)$ provides the following control:
$$  
\int_0^1 \Delta^{p-3}(u) du \leq C_{3-p}\cdot\Delta_*^{-(2-p)},  
$$
where $C_{3-p}=\frac{C(3-p)}{2-p} $.
Combining this with the previously established bound for $c_p$ and $|\Delta'(u)| \leq 2A$ yields the third term's upper bound $B_1\cdot\Delta_*^{-1}$, where  
$$
B_1= 2pA\cdot C_{p}\cdot C_{3-p} = \frac{2p^2AC^2(3-p)}{(p-1)(2-p)}.  
$$  

The fourth error term analysis parallels this approach. Starting with the integral  
$$  
\int_0^1 \frac{A u'(s')^2}{\Delta^3(u(s'))} ds' = c_p A \int_0^1 \Delta^{p-3}(u) du,  
$$  
we apply identical bounding techniques.
Substituting the normalization constant and gap integral estimates produces upper bound $B_2\cdot\Delta_*^{-1}$, where
$$
B_2 = A\cdot C_{p}\cdot C_{3-p} = \frac{pAC^2(3-p)}{(p-1)(2-p)}.
$$  
  
Combining all terms yields the total error bound:  
$$  
\eta(1) \leq  B_0\cdot\Delta_*^{-1}=\mathcal{O}(\Delta_*^{-1}),
$$
where the constant 
$$
B_0 = 2C_p + B_1+B_2,
$$
remains finite for $p \in (1,2)$.
This universal $\Delta_*^{-1}$ scaling demonstrates the adiabatic error's fundamental dependence on the minimal spectral gap.  
\qed
\vspace{0.8em}

This result indicates that, compared to the second-order dependence on the minimum spectral gap in \Cref{thm:error_estimation_wo_sf}, the upper bound of the error can be reduced to a first-order dependence on the minimum spectral gap by leveraging both the measure condition and the power-law condition. 
This demonstrates that when the spectral gap function satisfies the measure condition, an adaptive time interpolation function that fulfills the power-law condition can effectively reduce the error and enhance the precision of the adiabatic evolution.

\section{Optimality analysis}

In the previous section, we theoretically demonstrated that when the spectral gap function satisfies the measure condition, an adaptive time scheduling function obeying the power-law condition can effectively reduce the error and enhance the accuracy. 
This naturally raises a new question: are there alternative scheduling strategies that could further reduce the error?

To address this question, in this section we employ a variational approach to perform an optimality analysis. 
Ultimately, we will show that under certain conditions, the adaptive scheduling strategy based on the power-law condition is indeed optimal, meaning that no other choice of scheduling function can further reduce the error.

\subsection{Variational method}

Recalling \Cref{thm:general_schedule}, at a fixed time point $s = 1$, the total accumulated error in the evolution process depends on the time interpolation function in a functional manner.
This observation motivates the formulation of a functional optimization problem over the following function space:
\[
\left\{ u \in C^4[0,1] \,\Big|\, 
\begin{array}{l}
u: [0,1] \to [0,1], \ u(0)=0, \ u(1)=1
\\u'(0) = a \text{ fixed and}  \ u'(1) = b \text{ fixed.} \end{array}
\right\},
\]
Remarkably, the first and the second terms of error remain scheduling-independent constants for $u(s)$ in such function space. 
The optimization thus reduces to minimizing the following error functional:
\begin{equation}
\label{eq:loss_functional}
    \mathcal{I}[u] = \int_0^1 \mathcal{L}\left(s,u(s),u'(s),u''(s)\right) ds,
\end{equation}
with the following Lagrangian:
\begin{equation}
\label{eq:loss_l}
    \mathcal{L}(s,u,p,q) = \frac{|q|}{\Delta^2(u)} + \frac{A p^2}{\Delta^3(u)}.
\end{equation}
Therefore, we can use variational methods to explore the necessary condition for the minimum value to be satisfied.

\begin{thm}[First-order optimality condition]
\label{thm:optimality}\,\\
\noindent A scheduling function \( u(s) \) achieving local minimality for the error functional must satisfy the first-order variational Euler-Lagrange variational equation:
\begin{equation}
\label{eq:euler_lagrange}
\begin{aligned}
&
\frac{-2|u''(s)|\cdot\Delta'}{\Delta^3}
+
\frac{3A\cdot(u'(s))^2\cdot\Delta'}{\Delta^4}
-
\frac{2A\cdot u''(s)}{\Delta^3}
+
\frac{6\,\mathrm{sgn}(u''(s))\cdot(\Delta')^2\cdot(u'(s))^2}{\Delta^4}
\\
&
\quad\qquad
-
\frac{2\,\mathrm{sgn}(u''(s))\cdot\Delta''\cdot(u'(s))^2}{\Delta^3}
-
\frac{2\,\mathrm{sgn}(u''(s))\cdot\Delta'\cdot u''(s)}{\Delta^3} 
=
0,
\end{aligned}
\end{equation}
where we employ notations for short: \( \Delta = \Delta(u(s)) \), \( \Delta' = d\Delta/du \).
Here \( \mathrm{sgn}(\cdot) \) denotes the sign function and the equality holds in the $L^2$ sense.
\end{thm}

For simplicity, we present here the final result of the Euler–Lagrange equation directly. 
The detailed derivation can be found in Appendix A. 
The following development in this section will focus on rigorous investigation of this Euler-Lagrange equation \cref{eq:euler_lagrange}. 

\subsection{Necessity of nonlinear scheduling}

A natural question regarding the adaptive time interpolation strategy we introduced is: why adopt a power-law interpolation function in the first place? Would a simple linear evolution of the system (i.e., choosing $u(s) = s$) necessarily result in larger error?

To address this question, in this subsection we conduct a comparative analysis between the linear scheduling $u_0(s) = s$ and other nonlinear scheduling functions, such as those satisfying the power-law condition. 
This leads us to the following theorem.

\begin{thm}[Non-constant gap superiority]
\label{thm:nonconstant_gap}\,\\
\noindent For any non-constant spectral gap function \( \Delta \), there exists a nonlinear scheduling function \( u(s) \) with strictly lower error functional value than the linear schedule \( u_0(s) = s \):
\[
\mathcal{I}[u] < \mathcal{I}[u_0].
\]
\end{thm}

\begin{Proof}
We just need to notice that for $u(s)=s$ with $u'(s)=1$ and $u''(s)=0$, the left hand side of the first order variational equation \cref{eq:euler_lagrange} can be reduced into ${3A\Delta'(s)}/{\Delta^4(s)}$.
Under the assumption that $\Delta(u)$ is not a constant function, the aforementioned expression necessarily deviates from zero. 

This observation implies that the linear scheduling $u_0(s)=s$ fails to satisfy the Euler-Lagrange equation specified in \eqref{eq:euler_lagrange}. 
Consequently, the linear scheduling cannot correspond to a local minimum of the functional. 
This analytical result further demonstrates the existence of nonlinear scheduling functions capable of further reducing the value of the error functional, thereby providing  justification for pursuing non-uniform parameterization strategies in optimization frameworks.\qed

\end{Proof}
Through this theorem, we discover that if we do not choose scheduling, or in other words, adopt linear scheduling function \( u_0(s) = s \), then we can always reduce the error bound by reasonably choosing a nonlinear scheduling function as long as the spectral gap varies.
Specifically, the conventional linear scheduling scheme $u_0(s)=s$ fails to achieve minimal adiabatic error, as rigorously verified above, emphasizing the critical role of nonlinear scheduling.

\subsection{Optimality analysis}
Building upon the established necessity of nonlinear scheduling in the previous subsections, another natural inquiry arises: Are scheduling functions satisfying power-law constraints truly optimal within the function space of nonlinear parameterizations?
To address it, we now undertake a rigorous variational framework to characterize the necessary conditions for optimal scheduling functions. 
Let us first decompose the Lagrangian into two components:
$$
\mathcal{L}_1(s,u,p,q) = \frac{|q|}{\Delta^2(u)}, \quad \mathcal{L}_2(s,u,p,q) = \mathcal{L}_2(s,u,p) = \frac{p^2}{\Delta^3(u)}.
$$
This separation induces two independent error functionals:
$$
\mathcal{I}_1[u] = \int_0^1 \mathcal{L}_1(s,u,u',u'') ds, \quad \mathcal{I}_2[u] = \int_0^1 \mathcal{L}_2(s,u,u') ds,
$$
with the total error functional satisfying $\mathcal{I}[u] = \mathcal{I}_1[u] + A\mathcal{I}_2[u]$. The component-wise optimization will yield critical insights into scheduling dynamics. Actually, we have the following theorems for these two functionals.

\begin{thm}[Second-order derivative component optimality]
\label{thm:L1_optimal}\,\\
\noindent For linear gap profiles $\Delta(s) = \alpha s + \beta$, the power-law scheduling function with exponent $p = \frac{3}{2}$ satisfies the Euler-Lagrange equation derived from $\mathcal{L}_1$.
\end{thm}

\begin{Proof}
The first variation of $\mathcal{I}_1$ generates the Euler-Lagrange equation:
$$
\frac{d^2}{ds^2}\left(\frac{\partial \mathcal{L}_1}{\partial q}\right) - \frac{d}{ds}\left(\frac{\partial \mathcal{L}_1}{\partial p}\right) + \frac{\partial \mathcal{L}_1}{\partial u} = 0.
$$
Substituting $\mathcal{L}_1 = |u''|/\Delta^2(u)$ and simplifying yields:
$$
\frac{-2u''\Delta'}{\Delta^3} + \frac{6(\Delta')^2(u')^2}{\Delta^4} - \frac{2\Delta''(u')^2}{\Delta^3} - \frac{2\Delta' u''}{\Delta^3} = 0.
$$
For linear gaps ($\Delta'' \equiv 0$), this reduces to:
$$
\frac{2\Delta u'' - 3(u')^2\Delta'}{\Delta^4} = 0.
$$
The scheduling function $u(s)$ satisfying $u'(s) = c_p\Delta^{p}(u(s))$ with $p=\frac{3}{2}$ is exactly the solution.\qed
\end{Proof}
\begin{rem}
Moreover, \Cref{thm:L1_optimal} remains valid even when $\Delta(s)$ is piecewise linear, since in that case $\Delta''(s) = 0$ almost everywhere.
\end{rem}

This theorem formally establishes that temporal scheduling functions satisfying the power-law with exponent $p=\frac{3}{2}$ fulfill the necessary conditions for being local minima of $\mathcal{I}_1[u]$, provided the spectral gap function exhibits linear dependence. 
And for the second component, we have the following theorem.

\begin{thm}[First-order derivative component optimality]
\label{thm:L2_optimal}\,\\
\noindent The power-law scheduling function with exponent $p = \frac{3}{2}$ satisfies the Euler-Lagrange equation derived from the second component $\mathcal{L}_2$.
\end{thm}

\begin{Proof}
The first variation of $\mathcal{I}_2$ generates the Euler-Lagrange equation:
$$
\frac{d}{ds}\left(\frac{\partial \mathcal{L}_2}{\partial p}\right) = \frac{\partial \mathcal{L}_2}{\partial u}.
$$
Substituting $\mathcal{L}_2 = A(u')^2/\Delta^3(u)$ and simplifying reveals:
$$
2\Delta u'' = 3(u')^2\Delta',
$$
whose solutions correspond precisely to power-law schedules with $p = \frac{3}{2}$.\qed
\end{Proof}

This theorem demonstrates that scheduling functions satisfying the power-law constraint with exponent $p=3/2$ inherently fulfill the necessary conditions for local minima of $\mathcal{I}_2[u]$, without imposing additional constraints on the gap function.

Combining these two, we have the following theorem:
\begin{thm}[Optimality under linear gap]
\label{thm:opt_lgap}\,\\
\noindent Consider linear gap functions $\Delta(s) = \alpha s + \beta$. 
Assume the scheduling function $u(s)$ satisfies power condition with exponent $p = \frac{3}{2}$. 
Then $u(s)$ satisfies the Euler-Lagrange optimality condition \cref{eq:euler_lagrange}.
\end{thm}

\begin{Proof}
The verification proceeds through direct substitution analysis. Let $u(s)$ satisfy the power condition with $p = \frac{3}{2}$:
\begin{align*}
u'(s) &= c_p \Delta^{3/2}(u(s)), \\
u''(s) &= \frac{3}{2} c_p^2 \Delta^{2}(u(s)) \Delta'(u(s)).
\end{align*}
Beside, $\Delta''=0$ holds since $\Delta(s)$ is linear with $s$.
Then substituting all these into the Euler-Lagrange equation's transformed representation:
\begin{equation*}
\frac{1}{\Delta^4(u)}\Big[A + 2\,\mathrm{sgn}(u'')\Delta'\Big]\Big[3(u')^2\Delta' - 2\Delta\cdot u''\Big] - \frac{2\,\mathrm{sgn}(u'')\Delta''\cdot(u')^2}{\Delta^3} = 0.
\end{equation*}
It is trivial to reduce the entire expression to an identity $0 = 0$, confirming the theorem. In fact, this is also a direct corollary of \Cref{thm:L1_optimal} and \Cref{thm:L2_optimal}.\qed
\end{Proof}
The result in \Cref{thm:opt_lgap} directly follows as a corollary of \Cref{thm:L1_optimal} and \Cref{thm:L2_optimal}. 
Specifically, \Cref{thm:opt_lgap} establishes that when the spectral gap function exhibits (piecewise) linear dependence, time scheduling functions satisfying the power-law constraint with exponent \( p = \frac{3}{2} \) inherently fulfill the necessary conditions for local minima of the error functional with fixed boundary condition. 
This implies that the proposed power-law-constrained time scheduling strategy achieves conditional optimality under these spectral conditions, thereby providing a rigorous resolution to the fundamental question posed at the beginning of this section.

Within the variational framework developed in this section, we have systematically transformed the adiabatic error upper bound analysis into a functional optimization problem. 
Through rigorous application of the Euler-Lagrange formalism, we demonstrated that power-law-constrained time scheduling functions satisfy the necessary conditions for local extrema under specified spectral constraints. 
This theoretical guarantee substantiates the reliability and mathematical justification for employing such adaptive time scheduling strategies.

\section{Conclusion}
This work establishes a theoretical framework through the measure condition for spectral gaps and the power-law condition for time scheduling functions, reducing the adiabatic error's dependence on the minimal spectral gap from the conventional $\mathcal{O}(\Delta_*^{-2})$ scaling to an improved $\mathcal{O}(\Delta_*^{-1})$ scaling, as rigorously proven in \Cref{thm:error_estimation}.

The physical mechanism underlying this improvement from the adaptive sensing capability embedded in the spectral gap function. 
The system autonomously slows its evolution in regions of small spectral gaps to suppress error divergence, while accelerating through large-gap regions to maintain computational efficiency. 
This dual-regime strategy ensures that the accumulated evolution error remains significantly below the upper bound of uniform evolution protocols. 

Furthermore, through variational analysis, we provide optimality guarantees for such adaptive time scheduling functions under specified spectral conditions.
While spectral gap functions $\Delta(s)$ in generic Hamiltonian systems are typically nonlinear and often violate the measure condition, our framework still retains broader applicability through constructive approximations. 

For any smooth $\Delta(s)$ with a strictly positive lower bound $\Delta_*$, there exists a (piecewise) linear function $\Delta_l(s)$ satisfying $\Delta_l(s) \leq \Delta(s)$ with $\inf\Delta_l(s) = \Delta_*$. And then 
loosing the constrain between Hamiltonian and $\Delta$
and 
redefining the error metric $\eta(s; H(\cdot), \Delta(\cdot))$ in (\ref{eq:general_schedule_error}), we have
\begin{equation}
    \eta(s; H(\cdot), \Delta(\cdot)) \leq \eta(s; H(\cdot), \Delta_l(\cdot)).
\end{equation}
Our optimization framework still remains valid for $\eta(s; H(u(\cdot)), \Delta_l(u(\cdot)))$ because $\Delta_l(s)$ inherently satisfies the measure condition through piecewise linearity, power-law scheduling with $p=3/2$ maintains $\mathcal{O}(\Delta_{*}^{-1})$ scaling
\begin{equation}
    \eta(s; H(u(\cdot)), \Delta_l(u(\cdot))) \leq \mathcal{O}(\Delta_{*}^{-1}), 
\end{equation}
and variational analysis confirms that these functions satisfy necessary minimality conditions. 

The construction of $\Delta_l(s)$ proceeds as follows. 
We employ constant approximation near local minima, secant-line approximation in convex regions, and tangent-line approximation in concave regions, as discussed below \Cref{def:measure_condition}. 
This hybrid approach preserves the measure condition.
Thus, sufficiently smooth spectral gap functions also permit systematic construction of error-optimized scheduling functions.

While our framework advances adiabatic computation theory, some critical questions persist. 
For example, extending continuous optimization strategies to quantum circuit models requires additional analysis on the discrete level to maintain linear error scaling. 
Besides, for Hamiltonians with boundary cancellation condition, i.e., $H^{(k)}(0)=H^{(k)}(1)=0$ ($\forall k\geq1$), a candidate scheduling function 
\begin{equation}
    u(s) = c_e^{-1}\int_0^s \exp\left(-\frac{1}{t(1-t)}\right)\mathrm{d}t, \quad c_e = \int_0^1 \exp\left(-\frac{1}{t(1-t)}\right)\mathrm{d}t,
\end{equation}
requires rigorous analysis to determine its theoretical foundation and error suppression capabilities. 
Moreover, extending these principles to Hamiltonians of the form $H(s) = \sum_{i=0}^n u_i(s)H_i$ with multiple schedule functions still remains an open challenge requiring further works.

\section*{Acknowledgments}

DA acknowledges the support by the Quantum Science and Technology-National Science and Technology Major Project via Project 2024ZD0301900, and the Fundamental Research Funds for the Central Universities, Peking University.

\bibliographystyle{unsrt}
\bibliography{ref}

\begin{thebibliography}{10}

\bibitem{AlbashLidar2018}
Tameem Albash and Daniel~A. Lidar.
\newblock Adiabatic quantum computation.
\newblock {\em Rev. Mod. Phys.}, 90:015002, 2018.

\bibitem{HuyghebaertDeRaedt1990}
J.~Huyghebaert and H.~De~Raedt.
\newblock Product formula methods for time-dependent {S}chr\"{o}dinger problems.
\newblock {\em J. Phys. A}, 23(24):5777--5793, 1990.

\bibitem{WiebeBerryHoyerEtAl2010}
Nathan Wiebe, Dominic Berry, Peter H{\o}yer, and Barry~C Sanders.
\newblock Higher order decompositions of ordered operator exponentials.
\newblock {\em Journal of Physics A: Mathematical and Theoretical}, 43(6):065203, 1 2010.

\bibitem{PoulinQarrySommaEtAl2011}
David Poulin, Angie Qarry, Rolando Somma, and Frank Verstraete.
\newblock Quantum simulation of time-dependent {H}amiltonians and the convenient illusion of {H}ilbert space.
\newblock {\em Phys. Rev. Lett.}, 106(17):170501, 2011.

\bibitem{WeckerHastingsWiebeEtAl2015}
Dave Wecker, Matthew~B. Hastings, Nathan Wiebe, Bryan~K. Clark, Chetan Nayak, and Matthias Troyer.
\newblock Solving strongly correlated electron models on a quantum computer.
\newblock {\em Phys. Rev. A}, 92:062318, 2015.

\bibitem{LowWiebe2019}
Guang~Hao Low and Nathan Wiebe.
\newblock {H}amiltonian simulation in the interaction picture, 2019.

\bibitem{BerryChildsSuEtAl2020}
Dominic~W. Berry, Andrew~M. Childs, Yuan Su, Xin Wang, and Nathan Wiebe.
\newblock Time-dependent {H}amiltonian simulation with $l^{1}$-norm scaling.
\newblock {\em Quantum}, 4:254, 2020.

\bibitem{AnFangLin2021}
Dong An, Di~Fang, and Lin Lin.
\newblock Time-dependent unbounded {H}amiltonian simulation with vector norm scaling.
\newblock {\em Quantum}, 5:459, 2021.

\bibitem{Yi2021}
Changhao Yi.
\newblock Success of digital adiabatic simulation with large trotter step.
\newblock {\em Physical Review A}, 104:052603, Nov 2021.

\bibitem{Kocia2022digital}
Lucas~K. Kovalsky, Fernando~A. Calderon-Vargas, Matthew~D. Grace, Alicia~B. Magann, James~B. Larsen, Andrew~D. Baczewski, and Mohan Sarovar.
\newblock Self-healing of trotter error in digital adiabatic state preparation.
\newblock {\em Physical Review Letters}, 131:060602, Aug 2023.

\bibitem{AnCostaBerry2025}
Dong An, Pedro C.~S. Costa, and Dominic~W. Berry.
\newblock Large time-step discretisation of adiabatic quantum dynamics, 2025.

\bibitem{LuHuangAnEtAl2025}
Yangyu Lu, Yifei Huang, Dong An, Qi~Zhao, Dingshun Lv, and Xiao Yuan.
\newblock Digital adiabatic evolution is universally accurate, 2025.

\bibitem{BornFock1928}
Max Born and Vladimir Fock.
\newblock Beweis des adiabatensatzes.
\newblock {\em Z. Physik}, 51:165--180, 1928.

\bibitem{Kato1950}
Tosio Kato.
\newblock On the adiabatic theorem of quantum mechanics.
\newblock {\em Journal of the Physical Society of Japan}, 5(6):435, 1950.

\bibitem{Nenciu1993}
G.~Nenciu.
\newblock Linear adiabatic theory. exponential estimates.
\newblock {\em Comm. Math. Phys.}, 152:479--496, 1993.

\bibitem{AvronElgart1999}
Joseph~E. Avron and Alexander Elgart.
\newblock Adiabatic theorem without a gap condition.
\newblock {\em Communications in Mathematical Physics}, 203(2):445–463, June 1999.

\bibitem{HagedornJoye2002}
George~A. Hagedorn and Alain Joye.
\newblock Elementary exponential error estimates for the adiabatic approximation.
\newblock {\em Journal of Mathematical Analysis and Applications}, 267(1):235--246, 2002.

\bibitem{AmbainisRegev2006}
Andris Ambainis and Oded Regev.
\newblock An elementary proof of the quantum adiabatic theorem, 2006.

\bibitem{JansenRuskaiSeiler2007}
Sabine Jansen, Mary-Beth Ruskai, and Ruedi Seiler.
\newblock Bounds for the adiabatic approximation with applications to quantum computation.
\newblock {\em J. Math. Phys.}, 48(10):102111, 2007.

\bibitem{LidarRezakhaniHamma2009}
Daniel~A. Lidar, Ali~T. Rezakhani, and Alioscia Hamma.
\newblock Adiabatic approximation with exponential accuracy for many-body systems and quantum computation.
\newblock {\em Journal of Mathematical Physics}, 50(10), October 2009.

\bibitem{CheungHoyerWiebe2011}
Donny Cheung, Peter Høyer, and Nathan Wiebe.
\newblock Improved error bounds for the adiabatic approximation.
\newblock {\em Journal of Physics A: Mathematical and Theoretical}, 44(41):415302, September 2011.

\bibitem{ElgartHagedorn2012}
Alexander Elgart and George~A. Hagedorn.
\newblock A note on the switching adiabatic theorem.
\newblock {\em Journal of Mathematical Physics}, 53(10), September 2012.

\bibitem{GeMolnarCirac2016}
Y.~Ge, A.~Moln\'ar, and J.~I. Cirac.
\newblock Rapid adiabatic preparation of injective projected entangled pair states and gibbs states.
\newblock {\em Phys. Rev. Lett.}, 116:080503, 2016.

\bibitem{MozgunovLidar2022}
Evgeny Mozgunov and Daniel~A. Lidar.
\newblock Quantum adiabatic theorem for unbounded hamiltonians with a cutoff and its application to superconducting circuits.
\newblock {\em Philosophical Transactions of the Royal Society A: Mathematical, Physical and Engineering Sciences}, 381(2241), December 2022.

\bibitem{RolandCerf2002}
J{\'{e}}r{\'{e}}mie Roland and Nicolas~J. Cerf.
\newblock Quantum search by local adiabatic evolution.
\newblock {\em Phys. Rev. A}, 65(4):042308, 2002.

\bibitem{SubasiSommaOrsucci2019}
Yi{\u{g}}it Suba{\c{s}}{\i}, Rolando~D. Somma, and Davide Orsucci.
\newblock Quantum algorithms for systems of linear equations inspired by adiabatic quantum computing.
\newblock {\em Phys. Rev. Lett.}, 122:060504, 2019.

\bibitem{AnLin2022}
Dong An and Lin Lin.
\newblock Quantum linear system solver based on time-optimal adiabatic quantum computing and quantum approximate optimization algorithm.
\newblock {\em ACM Transactions on Quantum Computing}, 3(2), 2022.

\bibitem{JarretLackeyLiuWan2019}
Michael Jarret, Brad Lackey, Aike Liu, and Kianna Wan.
\newblock Quantum adiabatic optimization without heuristics, 2019.

\bibitem{RezakhaniKuoHammaEtAl2009}
A.~T. Rezakhani, W.-J. Kuo, A.~Hamma, D.~A. Lidar, and P.~Zanardi.
\newblock Quantum adiabatic brachistochrone.
\newblock {\em Phys. Rev. Lett.}, 103:080502, 2009.

\bibitem{RezakhaniAbastoLidarZanardi2010}
A.~T. Rezakhani, D.~F. Abasto, D.~A. Lidar, and P.~Zanardi.
\newblock Intrinsic geometry of quantum adiabatic evolution and quantum phase transitions.
\newblock {\em Physical Review A}, 82(1), July 2010.

\bibitem{Isermann2021}
Stefan Isermann.
\newblock On the optimal schedule of adiabatic quantum computing.
\newblock {\em Quantum Information Processing}, 20(9):300, 2021.

\bibitem{MatsuuraBuckSenicourtZaribafiyan2021}
Shunji Matsuura, Samantha Buck, Valentin Senicourt, and Arman Zaribafiyan.
\newblock Variationally scheduled quantum simulation.
\newblock {\em Physical Review A}, 103(5), May 2021.

\bibitem{WanKim2022}
Kianna Wan and Isaac~H. Kim.
\newblock Fast digital methods for adiabatic state preparation, 2022.

\bibitem{ShinguHatomura2025}
Yuta Shingu and Takuya Hatomura.
\newblock Geometrical scheduling of adiabatic control without information of energy spectra, 2025.

\bibitem{BraidaChakrabortyChaudhuriEtAl2025}
Arthur Braida, Shantanav Chakraborty, Alapan Chaudhuri, Joseph Cunningham, Rutvij Menavlikar, Leonardo Novo, and Jérémie Roland.
\newblock Unstructured adiabatic quantum optimization: Optimality with limitations.
\newblock {\em Quantum}, 9:1790, July 2025.

\bibitem{HanParkChoi2025}
Mancheon Han, Hyowon Park, and Sangkook Choi.
\newblock The constant speed schedule for adiabatic state preparation: Towards quadratic speedup without prior spectral knowledge, 2025.

\end{thebibliography}

\newpage

\section*{Appendxi A. Derivation of \Cref{thm:optimality}}
\label{appendx: thm4}

For the error functional defined in \cref{eq:loss_functional} with its corresponding Lagrangian \ref{eq:loss_l}, we derive the first-order necessary condition for optimality through direct application of the following Euler-Lagrange equation:
\begin{equation}
    \begin{aligned}
        &
        \frac{\partial \mathcal{L}}{\partial u}
        (s, u(s), u'(s), u''(s)) 
        -
        \frac{d}{ds}
        \left( 
        \frac{\partial \mathcal{L}}{\partial p}
        (s, u(s), u'(s), u''(s))
        \right)
        \\
        &
        \qquad\qquad
        +
        \frac{d^2}{ds^2}
        \left(
        \frac{\partial \mathcal{L}}{\partial q}
        (s, u(s), u'(s), u''(s))
        \right)
        = 
        0.
    \end{aligned}
    \label{eq:E-L-Equation_appendxi_2}
\end{equation}
Taking partial derivatives of the components of $\mathcal{L} $, we have:
\begin{align}
    \dfrac{\partial}{\partial u}
    \mathcal{L}(s,\,u,\,p,\,q)
    &=
    \Bigg(
    \frac{-2|q(s)|}{\Delta^3(u(s))} 
    + 
    \frac{-3A\cdot(p(s))^2}{\Delta^4(u(s))}
    \Bigg)
    \cdot
    \Delta'(u(s)),
    \\
    \dfrac{\partial}{\partial p}
    \mathcal{L}(s,\,u,\,p,\,q)
    &=
    \frac{2A\cdot p(s)}{\Delta^3(u(s))},
    \\
    \dfrac{\partial}{\partial q}
    \mathcal{L}(s,\,u,\,p,\,q)
    &=
    \frac{\mathrm{sgn}(q(s))}{\Delta^2(u(s))}.
\end{align}
Here, $\frac{\partial}{\partial q}\mathcal{L}$ is interpreted as the subgradient with respect to $q$. 
In particular, for the nonsmooth term $|q|$, we choose the subgradient value $\frac{\partial |q|}{\partial q}\big|_{q=0}=0$ at $q=0$.
And denote $\Delta '(u)=\frac{d}{du} \Delta (u) $. 
Next differentiating with respect to $s$, we have:
\begin{align*}
    \frac{d}{ds}
    \left( 
    \frac{\partial \mathcal{L}}{\partial p}
    (s, u(s), u'(s), u''(s))
    \right)
    =&
    \frac
    {-6A\cdot (u'(s))^2\cdot\Delta'(u(s))}
    {\Delta^4(u(s))}
    +
    \frac
    {2A\cdot u''(s)}
    {\Delta^3(u(s))},
    \\
    \frac{d}{ds}
    \left(
    \frac{\partial \mathcal{L}}{\partial q}
    (s, u(s), u'(s), u''(s))
    \right)
    =&
    \frac
    {-2\mathrm{sgn}(u''(s))\cdot\Delta'(u(s))\cdot u'(s)}
    {\Delta^3(u(s))},
    \\
    \frac{d^2}{ds^2}
    \left(
    \frac{\partial \mathcal{L}}{\partial q}
    (s, u(s), u'(s), u''(s))
    \right)
    =&
    \frac
    {6\mathrm{sgn}(u''(s))\cdot(\Delta'(u(s)))^2\cdot(u'(s))^2}
    {\Delta^4(u(s))}
    \\
    &+
    \frac
    {-2\mathrm{sgn}(u''(s))\cdot\Delta''(u(s))\cdot(u'(s))^2}{\Delta^3(u(s))}
    \\
    &+
    \frac
    {-2\mathrm{sgn}(u''(s))\cdot\Delta'(u(s))\cdot u''(s)}
    {\Delta^3(u(s))}.
\end{align*}
Based on the calculations above, we can convert \cref{eq:E-L-Equation_appendxi_2} into:
\begin{equation}
\label{eq:euler_lagrange_appendx2_thm4}
    \begin{aligned}
        &
        \frac{-2|u''(s)|\cdot\Delta'}{\Delta^3}
        +
        \frac{3A\cdot(u'(s))^2\cdot\Delta'}{\Delta^4}
        -
        \frac{2A\cdot u''(s)}{\Delta^3}
        +
        \frac{6\,\mathrm{sgn}(u''(s))\cdot(\Delta')^2\cdot(u'(s))^2}{\Delta^4}
        \\
        &
        \quad\qquad
        -
        \frac{2\,\mathrm{sgn}(u''(s))\cdot\Delta''\cdot(u'(s))^2}{\Delta^3}
        -
        \frac{2\,\mathrm{sgn}(u''(s))\cdot\Delta'\cdot u''(s)}{\Delta^3} 
        =
        0,
    \end{aligned}
\end{equation}
Here \( \Delta = \Delta(u(s)) \) and \( \Delta' = d\Delta/du \).
The equal sign here holds in the sense of $L^2$, and we complete the proof.

\end{document}